\documentclass[twocolumn,epsfile,prc,aps,showpacs]{revtex4}
\usepackage{graphicx}
\usepackage{epsfig}

\begin{document}

\title{On the structure of the  X(1835) baryonium
}

\author{ J.-P. Dedonder  \thanks{e-mail "dedonder@univ-paris-diderot.fr"}}
\affiliation{Laboratoire de Physique Nucl\'eaire et de Hautes \'Energies, Groupe Th\'eorie, IN2P3-CNRS, Universit\'es Pierre \& Marie Curie et Paris Diderot, 4 Place Jussieu, 75252 Paris, Cedex, France }

\author{ B. El-Bennich \thanks{e-mail "bennich@anl.gov"}}
\affiliation{Physics Division, Argonne National Laboratory, Argonne, IL 60439, USA}

\author{ B. Loiseau \thanks{e-mail "loiseau@lpnhe.in2p3.fr"}}
\affiliation{Laboratoire de Physique Nucl\'eaire et de Hautes \'Energies, Groupe Th\'eorie, IN2P3-CNRS, Universit\'es Pierre \& Marie Curie et Paris Diderot, 4 Place Jussieu, 75252 Paris, Cedex, France }

\author{S. Wycech\thanks{e-mail "wycech@fuw.edu.pl"}}
\affiliation{So{\l}tan Institute for Nuclear Studies,
Warsaw, Poland}

\begin{abstract}

The  measurement by the BES collaboration  of 
$J/\psi \rightarrow  \gamma  p {\bar p}$ decays
 indicates  an   enhancement at  the $p {\bar p}$
threshold. In another experiment  BES  finds a peak in the
invariant mass of  $\pi$  mesons produced in the possibly
related decay $J/\psi \rightarrow  \gamma  \pi^+ \pi^- \eta'$.
Using  a semi-phenomenological potential  model which  describes
all the $N {\bar N}$ scattering data, we show that the explanation
of both effects may be  given by a  broad quasi-bound state in the
spin and isospin singlet $S$ wave. The structure of the observed
peak is due to  an interference of this quasi-bound state with a
background amplitude and depends on the annihilation mechanism.
\end{abstract}

\pacs{12.39.Pn, 13.20Gd, 13.60.le,  13.75.Cs, 14.65Dw}
\maketitle

\section{Introduction \label{intro}}

The search for exotic states  in the  $N \bar{N}$ systems  has
been pursued  for a  few decades, but significant results have
only been obtained  recently. An indication of such states below
the $N \bar{N}$ threshold may be given by the scattering lengths
for a given spin and isospin state. However,  in scattering experiments, 
it is difficult to assess a  clear separation
of quantum states.
Measurements of the X-ray transitions in the antiproton hydrogen
atom  can select some partial waves  if  the fine structure of
atomic levels is resolved. Such resolution has been achieved for
the 1S states~\cite{aug99} and partly for the 2P
states~\cite{augnp99}. One can also use formation experiment
methods to reach specific states. In this way, an enhancement
close to the $ p\bar{p} $ threshold, has 
been observed by the BES Collaboration~\cite{bai03} in the radiative
decay

\begin{equation}
\label{1i}
 J/\psi \rightarrow \gamma p\bar{p}.
\end{equation}
On the other hand, a clear
threshold suppression is seen in the decay channel $
 J/\psi \rightarrow  \pi^0  p\bar{p}. $
To understand  better the nature of these $p\bar{p} $ states,  one
has to look directly into the subthreshold energy region. This may
be achieved   in the antiproton-deuteron or the antiproton-helium
reactions at  zero or  low energies. Such atomic experiments have
been performed, although  the fine structure resolution has not
been reached so far~\cite{got99,sch91}. Another way to look below
the threshold is the detection  of $N \bar{N}$ decay products.
Recently the  reaction

\begin{equation}
\label{3i}
 J/\psi \rightarrow  \gamma   \pi^+ \pi^- \eta'
\end{equation}
has been studied by the BES Collaboration~\cite{abl05}.
 This reaction  is attributed~\cite{abl05}  to an  intermediate
 $p \bar{p}$ configuration in the  $J^{PC}( p \bar{p} ) = 0^{-+}$ state
 which corresponds to spin singlet $S- $wave state. A peak in the invariant meson
 mass is observed, interpreted as a new baryon state, named  X(1835).
  The interpretation of the peak as
 a new X(1835) has been  questioned by the J\"{u}lich 
 group~\cite{jul06}. The latter  view is supported by our calculations,
 but we suggest the origin of the BES finding to differ from the
 possibilities presented  in Ref.~\cite{jul06}. It is argued  here
 that the peak  is due to an interference of a quasi-bound, isospin 0,
  $ N \bar{N}$ state with a background amplitude. The same
  quasi-bound state was found in Ref.~\cite{loi05} to be  responsible
  for the threshold
  enhancement in reaction (\ref{1i}).

The purpose of the present work  is to discuss  the  physics of
 $N \bar{N}$ states produced  in these $J/\psi$ decays and relate it
 to atomic experiments. In reaction  (\ref{1i}) only three  $p\bar{p}$ final states
  are possible, as a consequence of  the  $J^{PC}$
conservation. These differ by the internal angular momenta and
spins. Close to the $p \bar{p}$ threshold  a distinctly  different
behavior of scattering amplitudes is expected in different states.
A further selection of states is possible, but one has to rely on
the analyzes of the elastic and inelastic $N \bar{N}$ scattering
experiments. This has been studied in Ref.~\cite{loi05} within the
Paris potential model~\cite{par04,par99,par94,par82}, which is
also used in the present work.

The  final $p \bar{p}$ states allowed by  $P$ and $C$
conservation in the $\gamma p \bar{p}$ channel are specified in
Table I. These are denoted as $^{2S+1}L_{J}$ or
$^{2I+1,2S+1}L_{J}$, where $S,L$ and $J$ are the spin, angular
momentum and total momentum of the pair, respectively,  while $I$
denotes the isospin. A unified picture and a better specification
in the radiative decays is achieved, semi-quantitatively, with an
effective three-gluon exchange model~\cite{loi05}. This
description indicates the final $ \gamma p \bar{p}$ state to be
dominated by the $p \bar{p}$ ~$^{11}S_0$ partial wave. In this
wave the Paris potential generates a $52$~MeV broad quasi-bound
state at $4.8$~MeV below threshold. This state is named  $N
\bar{N}_S(1870)$.  A similar conclusion has been reached by the
J\"{u}lich group although the Bonn-J\"ulich potential does not
generate a bound state in the
 $p \bar{p}$~$^{11}S_0$ partial wave~\cite{jul06}.

Under the assumption that the $\pi^+,\ \pi^-$ and $\eta'$ are produced
 in relative $S$ waves, the
reaction (\ref{3i}), if attributed to an intermediate $p \bar{p}$
as suggested  by the BES group,  is even more restrictive than the
reaction (\ref{1i}). It allows only  one intermediate state the
 $p\bar{p}$ $^{1}S_0$, which coincides with the previous findings.
The presence of an intermediate $p \bar{p}$ state  in reaction (\ref{3i}) is
possible but not granted. We show below that a more consistent
interpretation is obtained with the dominance of the $N \bar N(1870)$
state which  is a mixture of $p\bar{p}$ and $n\bar{n}$ pairs.

The content of this work is as follows. Sec.~\ref{Sec1}
contains a  description of the final state $p \bar{p}$ interaction
and is included here for completeness. 
The  subthreshold $N \bar{N}$ scattering amplitude,
needed to describe reaction~(\ref{3i}), is defined in Sec.~\ref{Sec2}. Sec.~\ref{Sec3} gives the equation to be solved to calculate the amplitude of the meson formation through the intermediate $N \bar N$ interaction.  The results are
presented and discussed in Sec.~\ref{Sec4} together with some concluding remarks . 

\begin{table}
\caption{ The states  of the low-energy  $p \bar{p}$   pairs allowed
in the $ J/\psi \rightarrow \gamma p\bar{p}$ decays. The first
column gives the decay  modes to  the specified internal states of the 
 $p\bar{p} $ pair.  The $J^{PC}$ for the photon is $ 1^{--}$. The
second  column gives the $J^{PC}$ for the internal $p \bar{p}$ system,
the last column gives the relative angular momentum of the photon
vs. the pair. $ J^{PC} = 1^{--}$ for $ J/\psi $. \label{table1}  }
\begin{ruledtabular}
\begin{tabular}{lcc}
Decay mode                               &  $J^{PC}( p \bar{p} ) $ & relative l      \\
\hline
$\gamma  p \bar{p} (^1S_0) $               &  $0^{-+}$               &  1              \\
$\gamma  p \bar{p} (^3P_0) $               &  $ 0^{++}$              &  0             \\
$\gamma  p \bar{p} (^3P_1)$             &  $ 1^{++}$              &  0           
\end{tabular}
\end{ruledtabular}
\end{table}

\begin{figure}
\includegraphics[height=8.5cm,angle=90]{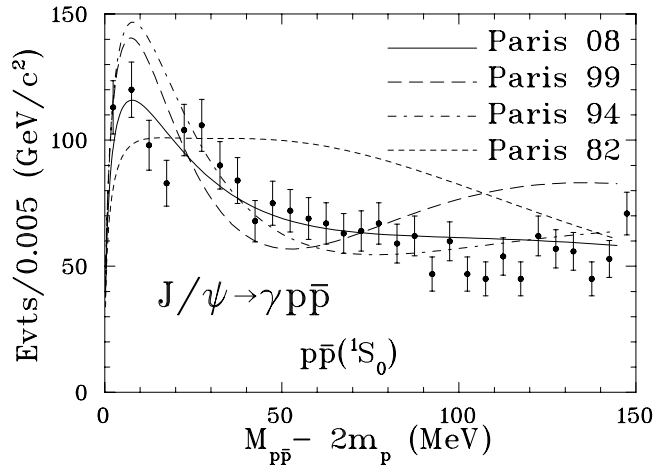}
\caption{ \label{fig1} The final state factor $q_f\left \vert
T_{if} \right \vert^2 $ for the $J/\psi$ decays  into $\gamma $ and $ p
\bar{p}$. The latest version of the Paris model generates a
quasi-bound state of $\Gamma = 52 $~MeV and $ 4.8 $~MeV binding
energy, and is the most consistent with the data.}
\end{figure}

\section{ Final state interactions \label{Sec1}}

 For any multichannel system at low energies, described by
an $S$ wave  $K$-matrix,
 the transition amplitude from an initial  channel $i$ to a final channel $f$ may be
described by

\begin{equation}
\label{1}
T_{if}= \frac{A_{if}  }{ 1   +   i  q_{f} A_{ff} },
\end{equation}
where $A_{if}  $ is a transition length,  $A_{ff}  $ is the
scattering length in the channel $f$ and $ q_{f} $ is the momentum
in this channel~\cite{pil67}. In the following,  channel $f$ is
understood to be the  $p \bar p$ channel. Within the same
formalism  the scattering amplitude in channel $f$ reads

\begin{equation}
\label{2}
T_{ff}= \frac{A_{ff}  }{ 1   +   i  q_{f} A_{ff} }.
\end{equation}
For $S$ waves at low energies $A_{if},A_{ff}$ are functions of $
q_{f}^2 $ and the main energy dependence of the amplitudes comes
from the denominators in Eqs. (\ref{1}) and (\ref{2}). With  large
values of Re $ A_{ff} > 0$  one may expect a bound (quasi-bound)
state. For large  Re $ A_{ff} < 0 $ a virtual-state is likely, but
one cannot determine these properties with absolute certainty
unless a method to extrapolate below the threshold exists. This is
particularly true in the  $p \bar{p}$ case where the absorptive
part Im~ $ A_{ff}$ is large because of the presence of many open
annihilation channels. Since  the final photon interactions are
believed to be  negligible,   the energy dependence observed in
the $J/\Psi \rightarrow \gamma  p \bar{p} $ decay  rate reflects
the energy dependence in  $q_f \left \vert T_{if}\right \vert^2$.

\begin{figure}
\includegraphics*[height=8.5cm]{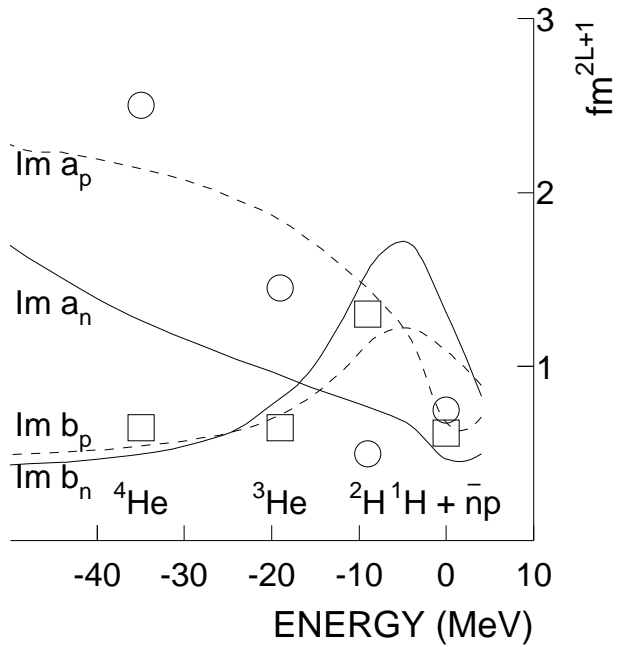}
 \caption{\label{fig2} The absorptive parts  of spin-isospin
averaged $N \bar{p} $ scattering amplitudes extracted
 from the atomic level widths in  H, De,
$^3$He and $^4$He $\bar{p}$~\cite{lea05}. Squares: $S$ waves and
circles: $P$ waves. The bottom scale indicates the energy below
threshold.
 The curves, calculated with the
 Paris 2008  potential, give the 
amplitudes separately:  $a_{n(p)}$ denote the $n \bar{p} $ or $p \bar{p} $ $S$-wave amplitudes, respectively and $b_{n(p)}$
the corresponding $P$-wave  amplitudes. The strong increase of absorption in the
$\bar{p}p$ $S$  wave is attributed mainly to the $^{11}S_0 $
state. }
\end{figure}

Practical calculations also indicate  an energy dependence in
$A_{ff}$ and the Watson approximation,  i.e., the constant $A_{if}$ is
not applicable in a broader energy range. One needs to use
Eq.~(\ref{1}) and  a weakly energy dependent formation amplitude
$A_{if}\sim 1/(1+q_f^2r_i^2)$ as explained in Ref.\cite{loi05},
where a best fit value $r_i=0.55$~fm was found.
 Figure~\ref{fig1}  displays sizable model dependence of   $q_f\left \vert T_{if}\right \vert^2 $
for the $^1S_0 $ calculated for four versions of the Paris
potential  model~\cite{par82,par94,par99,par04}. These versions
followed the increasing data basis which, for the most recent
case,  includes antineutron scattering and antiprotonic hydrogen
data. The threshold enhancement is attributed to a strong
attraction in this partial wave. It does not prove the existence
of  a quasi-bound state, but such a state is indeed generated by
the model in the $^{11}S_0 $ wave~\cite{par04}. There are
additional arguments to support this result which  follow from
light $\bar{p}$ atoms.
 The absorptive  amplitudes can be
extracted from the  atomic level widths. With the
 data from Refs.~\cite{got99,sch91} such an extraction  was described  in
Refs.~\cite{lea05} and~\cite{wyc07}. The data   allow   to obtain
only an isospin-spin average but as indicated in Fig.~\ref{fig2}, 
the existence of a quasi-bound state is consistent with the atomic
data. The increase of subthreshold absorption is also supported by
the atomic level widths in  heavy $\bar{p}$  atoms~\cite{wyc07}.
In addition to the broad $S-$wave state the Paris potentials
generate a narrow $^{33}P_1 $ quasi-bound state which arises in
Paris 08 and Paris 99 potentials. It gains some support from
widths in the antiprotonic deuterium  as indicated in
Fig.~\ref{fig2}.

 The procedure outlined above is
based on a simple form of the low-energy final state wave function
$\Psi_{p \bar{p}}$.  At large distances and for small $q_f r$, it becomes

\begin{equation} \label{4}
\Psi(r)_{p \bar{p}}  \sim 1 - \frac{T_{ff} \exp(iq_fr)}{r} \approx
[ 1-A_{ff} /r] \frac{1}{1+iq_fA_{ff} }.
\end{equation}
The right side of this relation  expands the wave  up to  $q_f^2$
terms. 
At shorter distance the wave function is no more directly related
to the scattering matrix and depends on details of the interaction
in  channel $f$.
 Integrated over an  unknown  transition potential $V_{if}$
it generates the formation amplitude   $A_{if} $ in the transition
amplitude $T_{if}$.  Eq.~(\ref{1}) was used with the Paris
\cite{loi05} and J\"ulich~\cite{jul05} potentials. These
potentials also generated the $A_{ff}$. Formulas (\ref{1}) or
(\ref{4}) are useful above the threshold but cannot be simply
extrapolated to  the subthreshold region. The difficulty is
related to the momentum  $ q_f = \sqrt{2\mu_{N \bar{N}}E_{N
\bar{N}} }$ where $\mu_{N \bar{N}} $ is the reduced mass. Above, 
the threshold $ E_{N \bar{N}}$ is the kinetic energy in the CM
system, below the threshold $ E_{N \bar{N}}$  is negative  and
$q_f$ becomes imaginary. The outgoing wave $ \exp(iq_fr)/r$
becomes $\exp(-\left \vert q_f \right \vert r)/r$. It damps strongly the
interaction  term in Eq.~(\ref{4}) and a more precise description
is necessary. We now turn to this point.

\begin{figure}
\includegraphics*[height=7.cm]{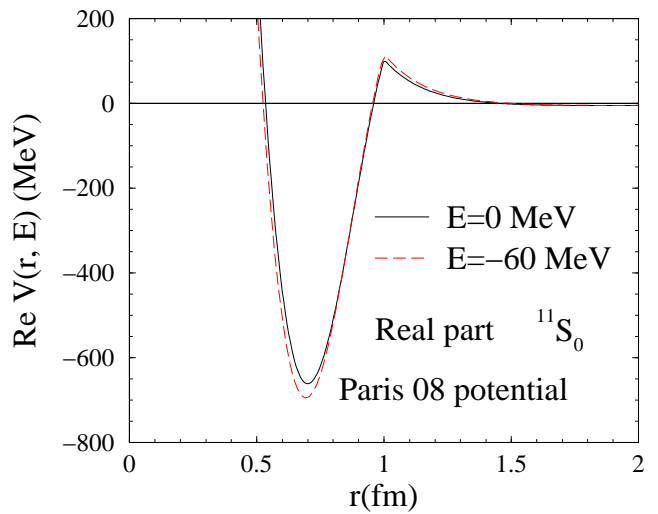}
 \caption{ \label{fig3}(Color online) The real Re~$V(r, E)$ potential
 for   $ N \bar{N} $  interactions  in the
$ ^{11}S_0 $ state. }
\end{figure}

\begin{figure}[ht]
\includegraphics*[height= 7 cm]{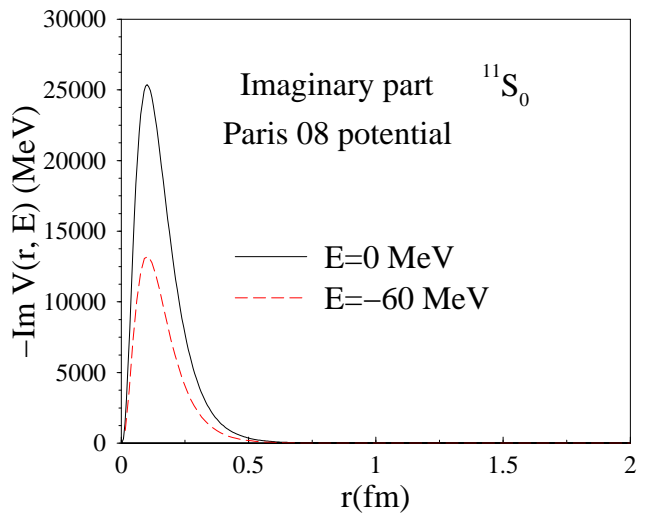}
 \caption{ \label{fig4}(Color online) The absorptive $-$Im~$V(r, E)$  potential
 for   $ N \bar{N} $  interactions  in the
$ ^{11}S_0 $ state. }
\end{figure}

\section {Off-shell $N\bar{N}$  interactions \label{Sec2}}

\begin{figure}[ht]
\includegraphics*[height=8.5 cm,angle=90]{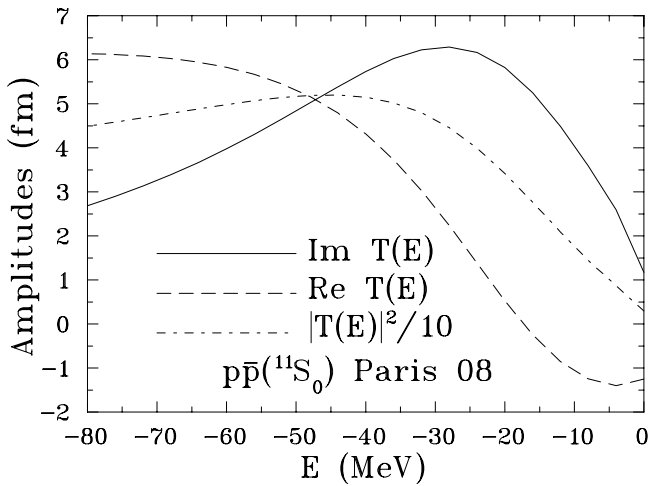}
 \caption{\label{fig5}
The  real  Re T(E) and imaginary  Im~T(E) parts  for   $ N \bar{N}
$ scattering amplitude in the $ ^{11}S_0 $ state. }
\end{figure}

 For further calculations one needs the off-shell extension of
the scattering amplitude in the energy as well as in the momentum
variables. The most general extension for $S$ waves  is given
by

\begin{equation}
\label{soff} 
f(k,E,k')= \frac{\mu_{N \bar{N}} }{2 \pi  }\int
~\psi_o(r,k)V _{N\bar{N}}(r,E)
\Psi^+(r,E,k')r^2~dr,
\end{equation}
where $\Psi^+(r,E,k')$ is the full outgoing wave calculated with
the regular free wave $ \psi_o(r,k')= \sin(rk')/(rk')$.
In this equation the momentum $k'$ is not related to the energy
$E$. The Fourier-Bessel double transform of $f(k,E,k')$ would
generate a nonlocal $\widetilde{f}(r,E, r')$ matrix in the
coordinate representation.  So involved  calculations  do not seem
necessary as the experimental data are rather crude. We resort to
a simpler procedure, standard in nuclear physics (for  an
application in the antiproton physics see Ref.~\cite{gre82}). The
subthreshold scattering amplitudes are calculated in terms of $T$
matrix defined in the coordinate representation by

\begin{equation}
\label{s1}  
\widetilde{T}(r,E)= \frac{\mu_{N \bar{N}} }{2 \pi  } ~
V_{N\bar{N}}(r,E)~ \frac{\Psi^+(r,E,k'(E)) }{\psi_o(r,k'(E))},
\end{equation}
with $k'(E)=\sqrt{2\mu_{N\bar{N}} E}$.
The $ \widetilde{T}(r,E)$ is a local equivalent of the nonlocal T matrix in
the sense that matrix elements in the $S$ waves fulfill the
relation $  f(k,E,k'(E)) = \int dr\ r^2\ 
\psi_o(r,k) \widetilde{T}(r,E)\psi_o(r,k'(E))$ valid in a narrow subthreshold
region where the last integral is convergent. 
The $V_{N\bar{N}}(r,E) $ is the recent Paris interaction potential
\cite{par04} which is used in the Schr\"odinger equation to
calculate $ \Psi^+(r,E,k'(E))$. The potentials used are plotted in Figs.~\ref{fig3} and \ref{fig4} and the resulting scattering amplitude is
given in Fig.~\ref{fig5}.

To describe the intermediate ${N \bar{N}} $  state we need also
the Fourier transform of $T(r,E)$

\begin{equation}
\label{s2a} 
T(\kappa,E) = \int  d
\textbf{r}~ \widetilde{T}(r,E)~\frac{sin(\kappa r)}{\kappa r}.
\end{equation}
In the next step, 
Eq.~(\ref{s2a}) is used at negative energies $ E= E_{N \bar{N}}$.
For positive energies this equation is not
practical due to zeros in the denominator that occur in $ \widetilde{T}(r,E)$ [see Eq.~(\ref{s1})]  at
multiplicities of $k'= \pi/ r$. One could nevertheless use it
for $k'< \pi~/ r_{max} $ where  $ r_{max}$ is the
distance at which the potential is cut-off. One has to set  $r_{max}\simeq 2
$~fm if one wants to extend the calculations  up to energies of $ \approx
20$~MeV above the $N \bar{N}$ threshold.
The normalization of $\widetilde{T}(r,E)$ by $\psi_o(r,k'(E))$ in Eq.~(\ref{s1}) insures the convergence of the integral [Eq.~(\ref{s2a})] below threshold.
In order to see the predictions of our model above threshold, we shall replace in Eq.~(\ref{s1}) $\psi_o(r,k'(E))$ by unity for $E\ge0$.

 The relevant on-shell $S$-wave   scattering amplitudes  given by

\begin{equation}
\label{s2} T(E)= \int  d \textbf{r} ~  \widetilde{T}(r,E),
\end{equation}
are normalized to the corresponding scattering lengths at the
threshold. The results for $T(E)$ are plotted in Fig.~\ref{fig5}.
One can notice a resonant behavior of the imaginary  part.
Roughly, the structure of this amplitude is dominated by the
weakly bound state $N\bar N_S(1870)$ in this wave. However, the location of the
bound state given as a pole in the complex energy plane ( Re $E=1871.7 $~MeV, 
$\Gamma=52 $~MeV)  corresponds neither  to the
maximum in the Im~$T(E)$  which occurs at $E \approx 1840$~MeV nor
to the maximum of  $ \pi^+ \pi^- \eta'$ invariant-mass
distribution that occurs at 1835~MeV. The interpretation of the
X(1835)  turns out to be more involved.

The potentials which generate this state are plotted in Figs.~\ref{fig3} and \ref{fig4}. The real potential ( Fig.~\ref{fig3})
contains a very weak attractive tail, a repulsive barrier, a strong
energy dependent attraction in the $0.5-1.0$~fm range  and a
repulsive core~\cite{par94}. These features, modified by the energy
dependent annihilation potential (Fig.~\ref{fig4}) and the proximity of the
threshold, generate a rather untypical $N \bar{N}$ scattering
matrix in the subthreshold energy region. The width of the
$^{11}S_0$  bound state indicates some energy dependence.
Moreover, the bound state form-factor displays strong enhancement
in the subthreshold region which is a typical phenomenon of the
subthreshold extrapolations. Altogether a strong enhancement of
 Im~$T$ is generated in the region well below the actual binding
energy. As discussed above this effect  finds support in the
widths of the $\bar{p} $-atom levels indicated  in
Fig.~\ref{fig2}.

\section{The intermediate $  p \bar{p} $  states \label{Sec3}}

 We assume  that the photon in reaction  (\ref{3i})
 is emitted before the
annihilation into mesons has taken place, as it happens in
reaction (\ref{1i}). A specific model for that process was
suggested in Ref.~\cite{loi05} but it will not be needed here. We
assume however that the formation of the  $ N \bar{N} $ pair is
described by a source function $F_{i,f}$ and the annihilation by
another function $F_{f,mes}$. In this way the effect of the
intermediate $ N\bar{N} $ interactions can be described by an
amplitude for the meson formation

\begin{equation}
\label{t1} T_{i,mes}  =   \int d\mathbf{p} \ d\mathbf{p'}~F_{i,f}(p)
 \ G(\textbf{p}, \textbf{p}', E_{ N \bar{N}} )~F_{f,mes}(p',Q),
\end{equation}
where $G(\textbf{p},\textbf{p}',E_{ N \bar{N}} )$ is the full Green's
function for the intermediate $ N\bar{N}$ system.
 The  form   assumed for the annihilation amplitude is

\begin{equation} 
\label{s5} 
F_{f,mes}(p',Q)=\left \langle \exp(-(\textbf{Q}-\textbf{p}')^2~r_f^2)\right\rangle,
\end{equation}
where the angular average over $\textbf{Q}$ is indicated by the
brackets. This choice is motivated by simple  model considerations
and the simplest possible assumption that the two $\pi$ mesons in
reaction (\ref{3i}) are correlated to the  $f_0(600)$ (also named
$\sigma$ meson). The mass of the latter is assumed to be 500~MeV
in our calculations. The relative momentum \footnote{Denoting  by $M_N$,  $m_\sigma$ and $m_{\eta'}$ the masses of the nucleon, $\sigma$ and $\eta'$ mesons, respectively, one has $\sqrt{m_\sigma^2+\textbf{Q}^2}+\sqrt{m_{\eta'}^2+\textbf{Q}^2}= 2M_N-\left \vert E_{N\bar N}\right \vert$.} of the final $\eta'$
and $\sigma$ mesons  is denoted by $ \textbf{Q }$ while the
Gaussian profile  comes from quark rearrangement models of
annihilation which operate Gaussian wave functions.

The Green's function in Eq.~(\ref{t1}) may be expressed in terms of
the free Green's function $G_o$ and the $ N \bar{N}$ scattering
amplitude $T$ as

\begin{equation}
\label{t2}
 G = G_o  + G _o ~T ~G_o.
\end{equation}
Now, with  the scattering amplitude defined by Eq.~(\ref{s1}) and
Eq.~(\ref{s2a})  one obtains

\begin{eqnarray}
\label{s3} 
T_{i,mes}& = & \int~ d\mathbf{p} ~ d\mathbf{p'}~
F_{i,f}(p) G_o(p,E_{ N \bar{N}}) ~ \left[
\delta(\mathbf{p}-\mathbf{p'})\nonumber \right.
\\ & & + ~\left. T(\left \vert \mathbf{p}-\mathbf{p'}\right \vert,E_{ N
\bar{N}})~G_o(p',E_{ N \bar{N}}) \right]~ F_{f,mes}(p').\nonumber\\&&
\end{eqnarray}
The first term in Eq.~(\ref{s3}) corresponds to a  background
amplitude with a non-interacting $N \bar{N}$ pair. The second one
describes intermediate state interactions. The Green's function is 
$G_o (p, E) = 4\pi/ [ (2\pi)^3 (q_f^2- p^2 )] $ and the
normalization is chosen such that  $T$ in Eq.~(\ref{s3}) has 
dimension of length. Let us notice that below the $ N\bar{N} $
threshold both $q_f^2$  and $G_o$ are negative. Below the
quasi-bound state  $T$ is attractive (negative) and the
interference  in Eq.~(\ref{s3}) becomes constructive. This effect
extends the peak structure to lower energies.  We assume the
formation amplitude to be described by

\begin{equation}
\label{s4} F_{i,f}(p)= \frac{1}{1+  p^2 r_{i}^2}
\end{equation}
with  the range parameter  $r_{i}= 0.55$~fm  determined before
from the final interactions above the threshold~\cite{loi05}. The
normalization is arbitrary. The angular integrations  in
Eq.~(\ref{s3}) generate an  amplitude which  depends only on
$\vert \textbf{Q}\vert$; that  is due to the momentum dependence of
the half-off shell $T$ matrix and to the absence of any  preferred
direction in the initial $ N\bar{N} $ state.

A semi-free parameter $r_{f}$ is related to the radius parameter in
the quark models for the nucleon and mesons. The range of allowed
$r_f$ values is limited. The upper limit $r_f \approx 0.55 $~fm is
obtained assuming the r.m.s. radii of the quark densities  to be
equal to the electromagnetic radii (0.8~fm for baryons and 0.6~fm
for mesons). A lower limit $r_f \approx 0.25 $~fm is obtained with
the radii used in $NN$  interaction models based on quark
approaches~\cite{lac02} and quark rearrangement models of $
N\bar{N} $ annihilation~\cite{gre84}. These rely on  r.m.s radii
in the range $0.5-0.6$~fm for baryons and $ 0.4- 0.6 $~fm for
mesons.

The last factor needed in this calculation involves the four body
phase space for  $J/\psi \rightarrow \gamma  \pi^+ \pi^- \eta' $.
We follow Ref.~\cite{pil67} to find

\begin{eqnarray}
&&\hspace{-0.9cm}dL_4(M_J^2; P_{\gamma}, P_{\pi^+}, P_{\pi^-}, P_{\eta'})\nonumber \\
&=& \frac{(M_J^2-S_M)}{(2\pi)^2 4 M_J^2} 
dL_3(S_M; P_{\pi^+}, P_{\pi^-}, P_{\eta'})d S_M,
\end{eqnarray}
where $M_J$ is the mass of $J/ \psi$ and  $dL_3 $
 is the invariant phase space for the three meson system of invariant mass squared
 $S_M$. The $dL_3 $ may
be found in Ref.~\cite{kal64} and it  generates only a weak energy
dependence. The full phase space is used but one finds a simple
approximation $ dL_4 \sim \varepsilon/ ( m_{\eta'} +2m_{\pi}
+\varepsilon)^2$ with $\varepsilon = \sqrt{S_M} - m_{\eta'} -2m_{\pi} $
to work well in the whole region of interest.

All together the spectral function representing the X(1835) is
given by
\begin{equation} \label{Sdecay} X_{S} =   \left \vert
T_{i,mes} \right \vert^2 ~dL_4/ dS_M.
\end{equation}

\section{Results and concluding remarks\label{Sec4}}

\begin{figure}
\includegraphics[height=8.5cm,angle=90]{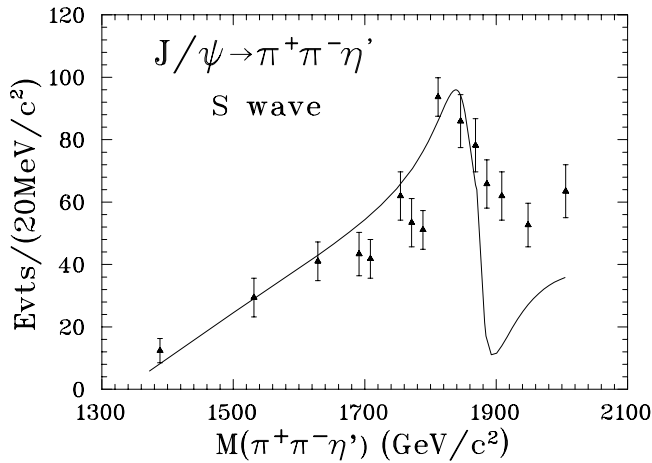}
 \caption{ \label{fig6}The spectral function X$_S$ representing the $X(1835)$ shape.
 Here the range parameter of the annihilation amplitude [Eq.~(\ref{s5})] is $r_f=0.45~fm$.
This $S$-wave contribution has been normalized to reproduce the data close to the $X(1835)$ peak. The experimental points are from Ref.~\cite{abl05}. Above the $N\bar N$ threshold the calculation is performed replacing $\psi_o(r,k'(E))$ by unity in Eq.~(\ref{s1}).}
\end{figure}

 The best description of the BES data is obtained
with $r_{f}\approx 0.4 - 0.5 $~fm, and the  shape of X(1835)
calculated in this way  is given  in Fig.~\ref{fig6}. The
intermediate state is  the isospin 0 state. The data are
reproduced fairly well despite the fact that the bound state
itself occurs at 1871.7~MeV, i.e., 4.8~MeV below the threshold.
This shape is determined by the interference effect of the two
terms in Eq.~(\ref{s3}) describing the decay process. Within
the Paris potential model and within a broad range of semi-free
$r_i, r_f$ parameters one finds no peak structure with the
intermediate $ p\bar{p}$ state. This result is consistent with the
observation that isospin 1 for the final mesons is not allowed and
the decay $ N\bar{N}(T=1) \rightarrow \pi^+ \pi^- \eta'$ is not
permitted by the isospin conservation~\cite{gre84}.

\begin{figure}
\includegraphics[height=8.5cm,angle=90]{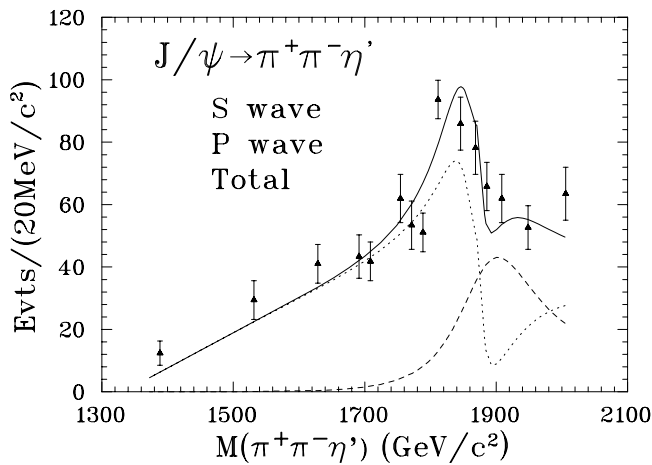}
 \caption{ \label{fig7}The spectral functions $X_S$ (calculated as for Fig.~\ref{fig6}) and  $X_P$ 
 (Eq.~(\ref{Pdecay}) with  $E_p=1900$~MeV and $\Gamma_p=200$~MeV) and their sum representing the X(1835) shape. Here The $S$- and $P$-wave contribution have been normalized to reproduced the data~\cite{abl05}.  }
\end{figure}

\emph{Other contributions, possible improvements.~~} Above the $
N\bar{N}$ threshold our estimation of $X_S$ from our model Eq.~(\ref{s3})
using Eq.~(\ref{s2a}) with  Eq.~(\ref{s1}) where  $\psi_o(r,k'(E))$ is replaced by unity 
[see our discussion in Sec.~\ref{Sec1} just below Eq.~(\ref{s2a})] generate a minimum (see Fig.~\ref{fig6}) that is
deeper than the minimum indicated by the data. Below we indicate
several possible explanations.

 \noindent $\bullet$
 The $J^{PC}$ conservation allows the $N\bar{N} $
pair in $^{11}S_0$ state to decay into the $f_0(600)~\eta'$ pair in a
relative $S$ wave. This  case has been discussed so far. In
addition, with the baryons in $^{13}P_1$ states another decay mode
to the $f_0(600)~\eta'$ pair is possible. It requires  the two final
mesons to be in the relative $P$-wave state. In the recent Paris
08~\cite{par04} as well as in the former Paris 99 potential~\cite{par99} a
close to threshold resonance is generated in the related
$^{13}P_1$ state. With the energy $E_P=1872$~MeV  and width
$\Gamma_P = 20 $~MeV it may contribute a spike to the spectralÄ
distribution in Fig.~\ref{fig6}. Since two different partial waves are
involved in the final states such a decay produces no interference
with the main mode and could contribute a term

\begin{equation}
\label{Pdecay} 
X_{P} = \left \vert \frac{ C_P ~Q }{ E_{N\bar{N}} - E_p +
i \Gamma_p /2} \right \vert^2 ~dL_4/dS_M
\end{equation}
to be added to  the main expression  for  $X_S $ given in Eq.~(\ref{Sdecay}).
The $X_P$  possibility is a speculative one and the relative
strength $C_P$ would be  very hard to predict. Also, with the
recent update of the Paris potential the position of $P$ wave
resonance is not generated at "the proper" position.

\noindent $ \bullet$ In a more complete study, outside the scope of the present work, one 
could extend our $S$-wave equations [Eqs.~(\ref{s1}), (\ref{s2a}) and  (\ref{s3})]
to the $P$-wave case. Here we illustrate in Fig.~\ref{fig7} the possible effect of an
effective $P$ wave represented by a resonant term given by Eq.~(\ref{Pdecay}).
It can be seen that such an effective resonance with $E_p=1900$~MeV and $\Gamma_p=200$~MeV can fill up
the above threshold dip of $X_S$.
 
 \noindent $ \bullet$ The off-shell extension in terms of
 Eq.~(\ref{s1}) cannot be fully trusted and the  procedure
of Eq.~(\ref{soff}) should be used.

\noindent $ \bullet$ The final state factor given by
 Eq.~(\ref{s5}) is perhaps too simple to be used above the
$N\bar{N}$ threshold. In some decay models, an energy dependent
phase factor $F_{mes}$ is expected ~\cite{gre84}. This would have
very limited effect below the threshold since the loop integrals -
over $G_o$ -   generate real functions. However,  above the
threshold the loop integral becomes complex  and the interference
pattern seen in figure (6) might be changed.

These effects go beyond the technique used in this paper.

\noindent $ \bullet$ To confirm experimentally a direct 
link between the $p \bar p$ system and the $X(1835)$, 
authors of Ref.~\cite{jul06} have suggested 
a search at the future GSI Facility for Antiproton and Ion Research
(FAIR) project using the proton antiproton detector array (PANDA) 
in  reactions such as $\bar p p \to \pi^+ \pi^- X$ and 
$X \to \pi^+ \pi^- \eta'$.
Another possible reaction would be
$\bar p p \to \gamma  X(1835)$. It could be performed with the PAX 
apparatus~\cite{PAX09}  with  $\sim$~50~MeV polarized 
antiprotons on polarized protons at CERN antiproton decelerator (AD) 
Ring.  
The shape of the $X(1835)$ could be tested by the photon energy 
distribution.
Of  special value  would be the comparison of two measurements  obtained
with the  parallel and anti-parallel  initial spin configurations. 
That could  give an  information
on the mechanism of the $X(1835)$ formation. 
In particular,  it would check the simple model
presented in Ref.~\cite{loi05},  where the initial state (in the J/Psi case 
the  intermediate) of the $p \bar p$ system is the spin triplet,  which,  
after  the emission of a magnetic photon, turns into the  final spin singlet.

  $In$ $summary$: it is shown that
the X(1835) structure can be generated by a conventional $N\bar{N}
$ potential model. Such a structure stems from  a broad and weakly
bound state, the $N\bar N_S(1870)$ that exists in the $^{11}S_0 $ wave. The existence of
a quasi-bound $S$-wave state receives an additional confirmation
from the level widths of antiprotonic atoms.

\vspace{.7 cm}

 Acknowledgements: We thank M. Lacombe for useful discussions. This research was performed in the framework of the IN2P3-Polish Laboratory Convention (collaboration N$^\circ$ 05-115).
One of us (S.W.) was supported by
EC 6-Th Program MRTN-CT-206-03502 (FLAVIA network).
This work was also supported in part  by the Department of Energy, Office of Nuclear Physics, Contract  No. DE-AC02-06CH11357.

\end{document}